\documentclass[table]{article}
\usepackage{authblk}
\usepackage{booktabs}
\usepackage{amsmath}
\usepackage{amssymb}
\usepackage{pgfplots}  
\usepackage{todonotes}
\usepackage[hyphens]{url}
\PassOptionsToPackage{table}{xcolor}
\pgfplotsset{compat=1.9}
\usepackage{caption}
\usepackage{subcaption}
\providecommand{\keywords}[1]{\textbf{\textit{Index terms---}} #1}

\definecolor{myblue}{HTML}{118ab2}
\definecolor{myred}{HTML}{ef476f}

\author[1]{Nikolaos Lykousas}
\author[2,3]{Constantinos Patsakis}

\affil[1]{Data Centric Services, Bucharest, Romania}
\affil[2]{Department of Informatics, University of Piraeus, 80 Karaoli \& Dimitriou str., 18534 Piraeus, Greece}
\affil[3]{Information Management Systems Institute of Athena Research Centre, Greece}

\usepackage[english]{babel}

\usepackage[letterpaper,top=2cm,bottom=2cm,left=3cm,right=3cm,marginparwidth=1.75cm]{geometry}

\usepackage{amsmath}
\usepackage{graphicx}
\usepackage[colorlinks=true, allcolors=blue]{hyperref}

\title{Tales from the Git: Automating the detection of secrets on code and assessing developers' passwords choices}

\begin{document}
\date{}
\maketitle

\begin{abstract}
Typical users are known to use and reuse weak passwords. Yet, as cybersecurity concerns continue to rise, understanding the password practices of software developers becomes increasingly important. In this work, we examine developers' passwords on public repositories. Our dedicated crawler collected millions of passwords from public GitHub repositories; however, our focus is on their unique characteristics. To this end, this is the first study investigating the developer traits in password selection across different programming languages and contexts, e.g. email and database. Despite the fact that developers may have carelessly leaked their code on public repositories, our findings indicate that they tend to use significantly more secure passwords, regardless of the underlying programming language and context. Nevertheless, when the context allows, they often resort to similar password selection criteria as typical users. The public availability of such information in a cleartext format indicates that there is still much room for improvement and that further targeted awareness campaigns are necessary.
\end{abstract}
\section{Introduction}

\section{Introduction}
One of the cornerstones of security is undoubtedly authentication, as it guarantees that only the prescribed entities are granted access to a given piece of information. Since credentials can be forgotten, intercepted, and stolen, there is a lot of work on biometrics and physically unclonable functions. The major advantage of these approaches is that one does not need to remember or bear anything and simply authenticates with what she is. Nevertheless, regardless of how seamless these modalities are, credentials are still the prevalent method to authenticate to a service.  

Contrary to their criticality, users, in general, do not make wise choices when selecting passwords. More precisely, users would choose guessable passwords, e.g. "password", celebrity names, bands, songs, dates, heroes from films and novels, easy to type passwords; e.g. \texttt{qwerty}, \texttt{qazxsw}, etc. This has been repeatedly reported and exploited, leading to the compromise of various systems and services. Therefore, service providers resort to applying password policies that would guarantee that the password would be difficult to guess, forcing the users to use upper and lower case characters, digits, special characters, be at least 8 characters long etc.  However, even this does not solve the problem. A typical user would resort to a simple variation of a typical password, e.g. \texttt{Passw0rd!}.

Regardless of the choice that a user makes about her password, a key issue is how they are stored. The reason is that one has to consider that the passwords must not be accessible from a malicious insider or a compromised host. Therefore, it is "prohibited" to store passwords in plaintext form. To allow for authentication without revealing the passwords, hashing and salting mechanisms are used and prevent, or at least significantly delay, attackers from performing offline attacks in case the password file has been leaked.

The above are well-known and widely studied problems. Going a step further, one has to consider the case of developers. A key difference here is that hard-coded credentials are distributed with their binaries and can greatly expose their organisation and infrastructure. There are numerous cases of such exposures, e.g., firmware~\footnote{\url{https://nakedsecurity.sophos.com/2021/01/06/zyxel-hardcoded-admin-password-found-patch-now/}}, security mechanisms\footnote{\url{https://thehackernews.com/2016/01/fortinet-firewall-password-hack.html}}, ICS controllers\footnote{\url{https://www.techtarget.com/searchsecurity/news/252442369/Yokogawa-Stardom-vulnerability-leaves-hardcoded-creds-in-ICS-controllers}}. Even more, the adoption of DevOps and Git has introduced another exposure as careless developers commit code where credentials, private keys, tokens and other possible sensitive information is exposed \cite{MeliMR19}. To detect credentials, tokens, and passwords, the mainstream approach is to look for high entropy strings or use regular expressions. Nevertheless, developers are not always so constrained in their writing. Their source code does not always comply with specific conventions and norms.

Based on the above, we investigate whether developers follow the same patterns that common users do when selecting passwords. Theoretically, since developers are more accustomed to computers and security, their choices should be significantly more secure and less predictable. To the best of our knowledge, no concrete study is exploring this problem. Merely asking people working on computers to select passwords triggers a bias as they feel that they are tested on this aspect and know how to properly address it. However, this does not necessarily mean that they would use the same criteria when they are not being monitored for this aspect and freely select them for the comfort of their daily work.
To answer the above questions, we leverage a large-scale dataset, the biggest available to date, containing more than 2 million secrets that were committed in public repositories on GitHub. This dataset contains labels according to programming language, type of secret (e.g., token, password), usage (e.g., mail, database) etc. Based on this dataset, we provide a thorough analysis of the passwords that developers actively use to detect patterns.

In the next section, we provide a brief overview of the related work. Then, we describe our data collection methodology and our dataset. Afterwards, in Section \ref{sec:methodology}, we outline our credential extraction methodology. In Section \ref{sec:analysis}, we analyse the dataset and discuss the structure and emerging patterns. Finally, the article concludes by summarising our contributions, limitations of our work, and possible extensions for future work.

\section{Related work}
\label{sec:related}
As already discussed, users often choose predictable passwords for their services \cite{malone2012investigating,von2013survival,JakobssonD12}. Some studies suggest that people with a computer science background select stronger passwords than others \cite{2508859.2516726}. However, users tend to have misconceptions regarding password generation. For instance, the interviews conducted by Ur et al. \cite{UrNBSSBCC15} uncovered several of them, e.g. adding a symbol or using a 'non-public' date or name would lead to a secure password. Most of these misconceptions can be attributed to the lack of users understanding of either the common patterns in which people think or the capacity to brute-force variations of known keywords for an individual. In fact, the passwords reflect several traits of ourselves and our culture \cite{alsabah2018your,wang2019birthday}.

From the above, one can derive several lines of research to protect users, as they still seem reluctant to use, e.g. password managers \cite{mayer2022users}. In fact, user passwords are shown to adhere to specific models, e.g. Zipf-like models \cite{7961213}. One way to make users use strong passwords is to prevent them from using weak ones by enforcing specific policies on the characters used in the password, e.g. mandating the use of longer passwords, use of letters (both small and capital), digits, and symbols. Despite the fact that this sounds like a valid policy, users often end up using a password of the form \texttt{Passw0rd!} or \texttt{Pa\$\$w0rd}, which cannot be considered much more secure. Therefore, one research line is to measure the strength of passwords without context, trying to determine the entropy of the string, the length, the complexity and length of the underlying patterns, the position of the characters on the keyboard etc. There are various approaches \cite{3025453.3026050,Wheeler16,melicher2016fast,HuZG20a,HoushmandA12} with varying accuracy \cite{golla2018accuracy} leveraging among others probabilistic context-free grammars, Markov models, and neural networks.

The fact that users tend to use a limited set of rules to generate passwords \cite{JakobssonD12} makes them guessable \cite{2976749.2978339}, with users often failing to understand the reasons why \cite{2858036.2858546}. The guessability of user passwords has emerged as another research line that tries to quantify this extent. To this end, researchers exploit  data-driven methods to assess how similar passwords are with leaked passwords \cite{melicher2016fast,pal2019beyond,hitaj2019passgan,xia2019genpass,pasquini2021improving} but also to train password crackers \cite{di2022revenge} and to prevent such attacks \cite{wang2022segments}.

Nevertheless, developers are not typical users, nor just savvy. They develop software solutions and regulations, such as the General Data Protection Regulation (GDPR) in the European Union and the California Consumer Privacy Act (CCPA) in the United States force them to apply secure policies on the collection, processing, and handling of personal data. Nevertheless, their policies are not the best. For instance, they do not store passwords \cite{3134082,naiakshina2020conducting} properly, misconfigure the underlying infrastructure \cite{dietrich2018investigating}, do not sanitise the input properly \cite{braz2021don}.

The wide adoption of DevOps pipelines from the industry and the use of Git has given rise to another issue: developers often make public commits with their code containing secrets and passwords. One of the first studies is that of Meli et al. \cite{MeliMR19}, who harvested thousands of API keys, private keys, and secrets from public GitHub repositories. This has sparked the development of targeted crawlers for code repositories and source code scanners to extract secrets and passwords \cite{saha2020secrets,10063545}. 

\section{Data Collection}
\label{sec:collection}
To collect hardcoded credentials from GitHub, we employ a method similar to \cite{MeliMR19} and use a set of targeted queries to collect candidate files possibly containing various types of secrets. Several of these queries were derived from the authentication-related code snippets provided by Feng et al. \cite{feng2022automated}. Moreover, we defined additional queries capturing authentication contexts not considered in \cite{feng2022automated}, e.g. login automation using Selenium Webdriver. Provided that at the time of wiring, GitHub API allowed only scoped code searches (specific repository/user/organization/etc.), we crafted a crawler leveraging the search interface of \url{http://github.com} instead of the API, and accordingly collected all the returned results. The dataset was collected over 6 months, from May to Oct. 2021. In total, approximately 30M files were collected. Our queries targeted a wide range of programming languages and configuration files, including: \texttt{C}, \texttt{C\#}, \texttt{C++}, \texttt{Classic ASP}, \texttt{Dart}, \texttt{Go}, \texttt{INI}, \texttt{JSON}, \texttt{Java}, \texttt{Java Properties}, \texttt{Java Server Pages}, \texttt{JavaScript}, \texttt{Jupyter Notebook}, \texttt{Kotlin}, \texttt{PHP}, \texttt{Python}, \texttt{R}, \texttt{Ruby}, \texttt{Scala}, \texttt{Shell}, \texttt{Swift}, \texttt{TSX}, \texttt{Text}, \texttt{TypeScript}, \texttt{Visual Basic .NET}, \texttt{Vue}, \texttt{XML}, and \texttt{YAML}.

\section{Credential Extraction}
\label{sec:methodology}
The inherent diversity of passwords makes detection approaches based on regular expressions and entropy-related heuristics ineffective \cite{feng2022automated}. Naturally, passwords lack a well-defined structure and frequently do not comply with strong password policies (which would ensure their distinctiveness compared to regular strings found in source code). As such, to effectively extract credentials from our dataset, we leverage methods from the PassFinder model for password leakage detection from source code presented in \cite{feng2022automated} while introducing several improvements considering the broader set of target languages we considered, compared to Feng et al. \cite{feng2022automated} More precisely, PassFinder employs an ensemble of two text convolution neural network models, namely the \textbf{Context Model} and the \textbf{Password Model}. The Context Model is trained to classify source code snippets surrounding specific seed elements (including names of methods, variables, constants, etc.) that can be potentially relevant for various authentication contexts. To establish the optimal number of lines of these snippets, Feng et al. performed a series of experiments and set the context window size at 6, which provides a good enough representation of the context in which these seed keywords appear. The Password Model is trained to classify strings extracted for the code snippets previously identified as being potential passwords, machine-generated secrets (e.g. API keys, JWT tokens, etc.), or ordinary strings.
To enhance the performance and accuracy, we made a series of improvements, including introducing a series of innovative features for the Password Model to assess the degree of human memorability for strings, drawing upon the research presented in Casino et al. \cite{casino2021intercepting}, augmenting the candidate password extraction from source code with an extensive set of regular expressions capturing cases where passwords are not wrapped in quotes (e.g. within connection strings, URLs, comments, etc.), as well as expanding the datasets used to train its two components. Nevertheless, these improvements fall outside the scope of the current work, and thus, a detailed exploration of their impact will be reserved for future research. Finally, we also altered the architecture neural network tasked with modelling the context semantics, i.e. the Context Model, to classify code snippets belonging to the following categories of credentials:
\begin{itemize}
\item  \textbf{Database}: Credentials used to authenticate connections to various data stores, including MySQL, Microsoft SQL Server, PostgreSQL, Oracle, MongoDB, etc.
\item  \textbf{Mail}: Credentials used to authenticate connections to mail servers, including SMTP, IMAP, and POP3 protocols.
\item  \textbf{Automation}: Strings submitted to log-in web forms using Browser Automation libraries, such as Selenium.
\item  \textbf{Web Service}: Credentials used to perform HTTP basic authentication, proxy authentication, as well as being part of HTTP request bodies or parameters.
\item  \textbf{Other}: Authentication-related code not fitting the previous categories.
\end{itemize}
To assess the performance of both components in our enhanced version of PassFinder, we manually examined the results in a random sample of 2,000 authentication snippets, which included code in languages not considered in \cite{feng2022automated}. The performance of our Password Model was better compared to Feng et al. ($Macro-F_1$ score 98.1\% vs 96.79\%), while the performance of the augmented Context Model in terms of credential category classification achieved a $Macro-F_1$ score of 88.7\%. It is important to note that we do not consider strings containing non-ASCII characters as potential passwords.

\section{Dataset}
\label{sec:dataset}
Next, we present the password dataset extracted by applying the credential extraction approach described in Section~\ref{sec:methodology} to the collected source code files (see Section~\ref{sec:collection}), as well as the dataset of leaked passwords we use for comparison.

\subsection{Hardcoded credentials on Github}
In total, we extracted 2,093,488 unique strings that were classified as passwords. Note that, for candidate passwords, we considered strings with a length of at least 5. From these, 425,071 were classified as machine-generated; that is API keys, tokens, encrypted strings, and  hashes, and 280,750 were included in wordlists of default credentials for various platforms\footnote{\url{https://github.com/danielmiessler/SecLists}}. To perform a fair comparison with the predominantly human-generated passwords in leaked credential databases, typically used for authenticating online accounts, we excluded these strings, resulting in a total dataset of 1,387,667 hardcoded passwords by 294,975 developers. In our dataset, the 78.95\% of developers (232,890) leaked a single password, while the maximum number of unique passwords associated with a single developer was 1,718, indicating their utilisation in a unit testing context.
Note that we do not examine password reuse across developers in this study as many developers may have used the same credentials for the same service multiple times, shared them within a team, or added them in multiple unit tests. Thus, we focus on the 439,204 unique passwords, which constitute 31.65\% of the human generated passwords in our dataset. 
For the rest of this paper, we refer to this dataset as \texttt{Developers}. Note that, given the large scale of our dataset, there might be instances of machine-generated passwords that could not be filtered out using the employed heuristics. Nevertheless, we expect that their impact on our comparison experiments will be negligible. 

\subsection{Leaked passwords}
For leaked user passwords, we use the so-called \texttt{RockYou2021} dataset. The dataset is a comprehension of previously leaked databases and contains 8.4 billion entries of passwords. The dataset was published in RaidForums, which has now been seized  by the U.S. Department of Justice (DOJ)\footnote{\url{https://www.justice.gov/usao-edva/pr/us-leads-seizure-one-world-s-largest-hacker-forums-and-arrests-administrator}} as the DOJ notes that:\begin{quote}
    \textit{Members could also earn credits through other means, such as by posting instructions on how to commit certain illegal acts.}
\end{quote}
While researchers claim that password choices do not vary greatly over time, we opted for this dataset as beyond its size and variety, it is expected to reflect the latest password policies on the selection of user passwords. For tractability reasons, in our experiments, we consider a random sample of 10M unique passwords.

\section{Password Analysis and Comparison}
\label{sec:analysis}

\begin{figure*}[th]
   \begin{subfigure}[t]{0.475\columnwidth}
	\centering
      \includegraphics[width=\columnwidth]{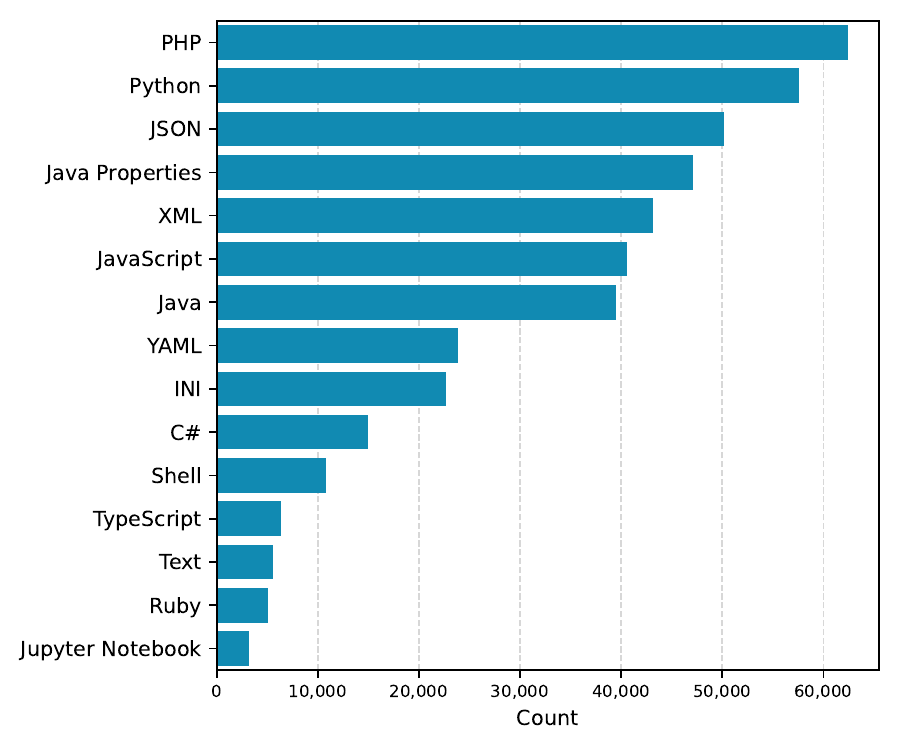}
      \caption{Top-15 Languages}
      \label{fig:top15_cnt}
   \end{subfigure}
   \hfill
   \begin{subfigure}[t]{0.475\columnwidth}
	\centering
	\includegraphics[width=\columnwidth]{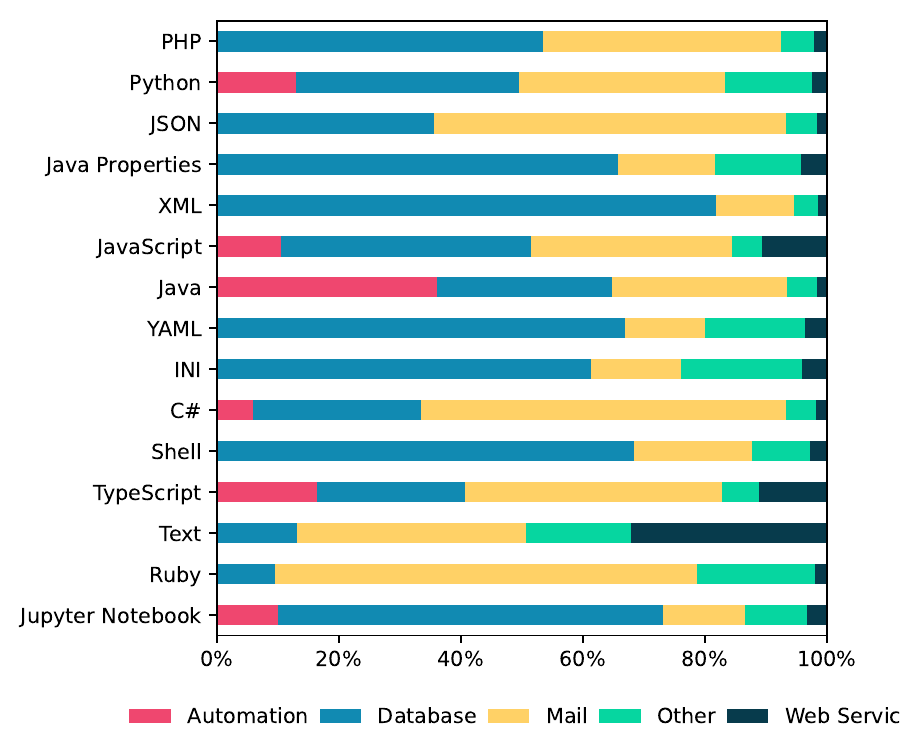}
    \caption{Credential Types}	
    \label{fig:top15_types}
	\end{subfigure}
  \caption{Number of passwords and proportions of different credential types for the Top-15 programming languages in our dataset.}
\end{figure*}

\begin{figure*}
   \begin{subfigure}[ht]{0.475\columnwidth}
	\centering
      \includegraphics[width=\columnwidth]{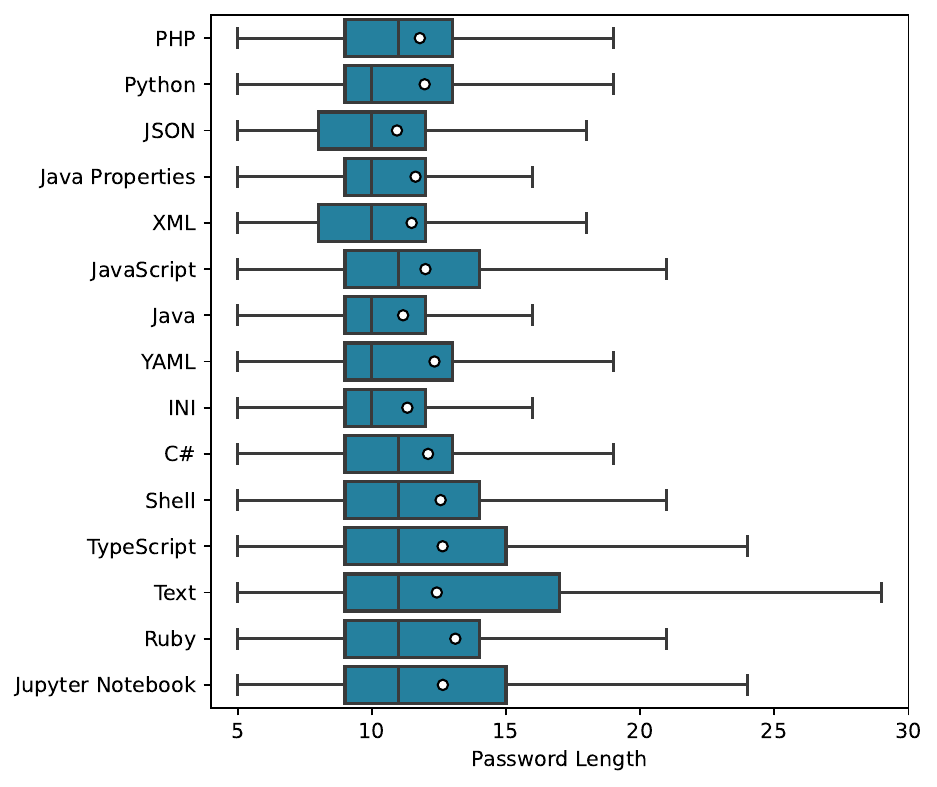}
      \caption{Lengths by language}
      \label{fig:password_len_by_language}
   \end{subfigure}
   \hfill
   \begin{subfigure}[ht]{0.475\columnwidth}
	\centering
	\includegraphics[width=\columnwidth]{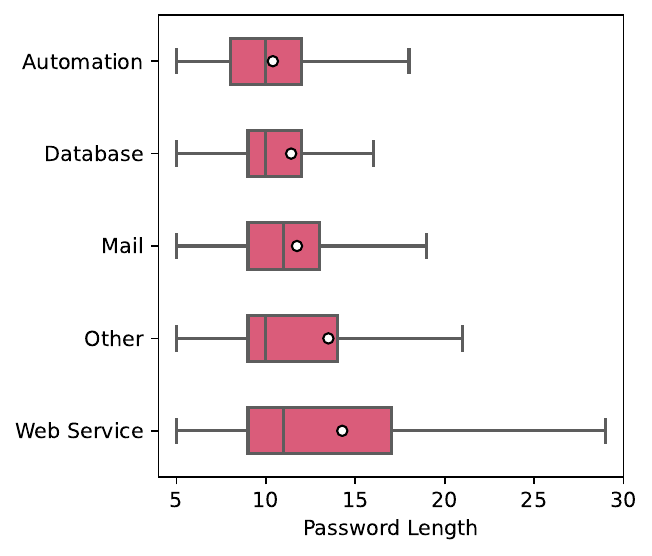}
    \caption{Lengths by type}	
    \label{fig:password_len_by_type}
	\end{subfigure}
  \caption{Password length distribution for the Top-15 languages and the 5 credential categories. The circular white markers represent the mean length per group.}
  \label{fig:password_len_box}
\end{figure*}

In Figure \ref{fig:top15_cnt}, the Top-15 languages in the \texttt{Developers} dataset are illustrated. We observe that although the majority of most popular programming languages \cite{diakopoulos2015interactive,cass2020top} is present, a significant number of credentials is contained in configuration files in standard formats, including JSON, Java Properties, YAML, and INI. Moreover, we notice a non-trivial number of passwords in Text files, meaning that developers can leak authentication snippets and credentials inside plaintext files, commonly used for note-taking purposes. The distribution of passwords in the different categories considered in Section~\ref{sec:methodology} are illustrated in Figure~\ref{fig:top15_types}. We observe that `Database' and `Mail' are the most prominent categories across all the top languages, while passwords in the `Automation' category do not appear in configuration files. Nevertheless, please note that these distributions are tightly related to the queries we performed for collecting the studied dataset, and as such, it is important to acknowledge the potential bias introduced by our specific choice of queries, which while capturing a representative sample of various password categories, may not provide a comprehensive view of all possible scenarios.

Next, we study the distribution of password lengths across the Top-15 programming languages and the 5 credential types in our dataset. Accordingly, we plot these distributions in Figure~\ref{fig:password_len_box}. In Figure~\ref{fig:password_len_by_language} we observe that the majority of passwords have relatively similar median lengths, ranging from 9 to 11 characters. Notably, the TypeScript language stands out with a higher mean length of 13.12 characters, while INI and JSON languages exhibit slightly shorter average password lengths at 10.40 and 10.46 characters, respectively. Figure~\ref{fig:password_len_by_type} focuses on the distribution of password lengths for each credential type. Here, we observe a more noticeable variation in password lengths across different types. `Web Service' and `Other' credential types exhibit the longest mean password lengths at 14.27 and 13.49 characters, respectively. In contrast, `Automation' credentials have a shorter mean length of 10.39 characters, while the `Database' and `Mail' types show moderately longer average password lengths, with means of 11.41 and 11.74 characters, respectively. These differences in password lengths across credential types suggest that the context in which a password is used may play a more substantial role in determining its length, with some authentication contexts requiring or encouraging the use of longer, and potentially more secure, passwords.

\begin{figure}[th]
\centering
	\includegraphics[width=.6\columnwidth]{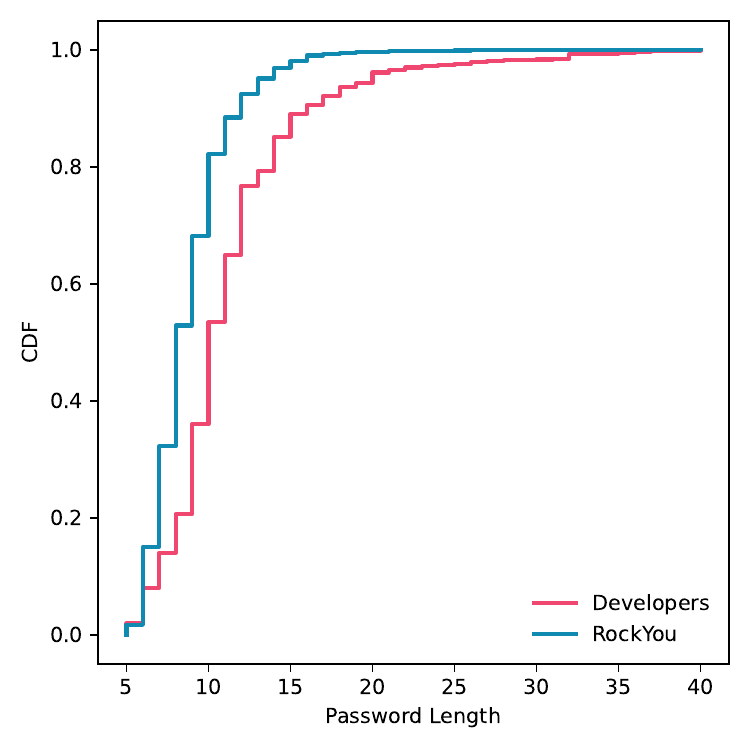}
    \caption{CDF of password lengths, trimmed at length 40 for readability, representing $>99\%$ of total passwords in both datasets.}
    \label{fig:cdf_pass_len}
\end{figure}

We proceed to perform a comparison between the characteristics of the passwords in \texttt{Developers} and \texttt{RockYou} datasets. In Figure~\ref{fig:cdf_pass_len} we plot the cumulative distribution function (CDF) for the password lengths for each dataset. A CDF displays the probability that a random variable will take a value less than or equal to a specific value. In this context, the CDFs illustrate the proportion of passwords with lengths less than or equal to a given length for both datasets. It is evident that developers use significantly longer passwords compared to typical users. For example, the probability of observing a password of length 10 or shorter is 53.46\% for the \texttt{Developers}, whereas the same probability for \texttt{RockYou} is significantly higher at 82.27\%. This trend continues for longer password lengths as well, with the developers' passwords consistently exhibiting lower cumulative probabilities at each length compared to normal users.

\begin{figure*}[th]
\centering
   \begin{subfigure}[hb]{0.475\columnwidth}
	\centering
      \includegraphics[width=\columnwidth]{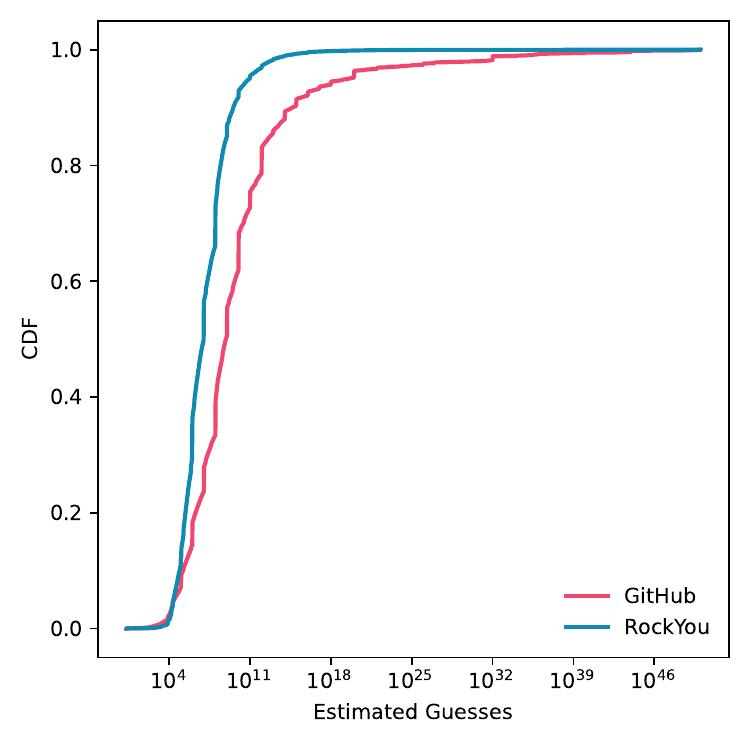}
      \caption{CDF of the estimated number of guesses}
      \label{fig:cdf_zx}
   \end{subfigure}
   \begin{subfigure}[hb]{0.475\columnwidth}
	\centering
	\includegraphics[width=\columnwidth]{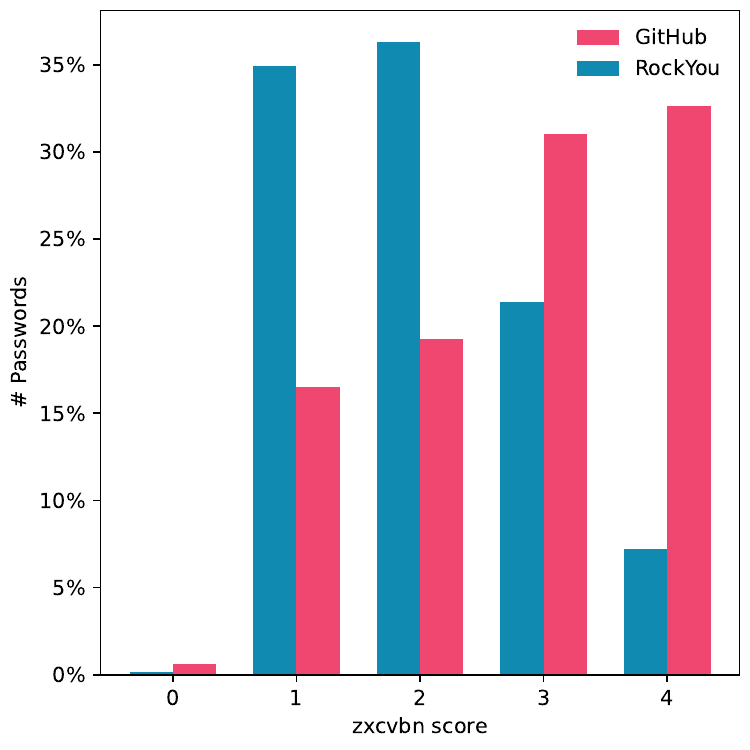}
    \caption{Scores}	
    \label{fig:score_zx}
	\end{subfigure}
  \caption{Comparison of \texttt{zxcvbn} metrics for both datasets.}
\end{figure*}

To further investigate the differences in password strength between the \texttt{Developers} and \texttt{RockYou} datasets we employ the well-known \texttt{zxcvbn}~\cite{Wheeler16}. Zxcvbn is a widely-used password strength estimator that evaluates the strength of passwords by estimating the number of guesses an attacker would require to crack them, how long it would take, and by providing a password score ranging from 0 (weakest) to 4 (strongest). To this end, we plot the CDFs of the estimated number of guesses for both datasets in Figure~\ref{fig:cdf_zx}. We observe that it follows the same trend as the password lengths (Figure~\ref{fig:cdf_pass_len}), indicating that the developers' passwords are generally stronger and more resistant to brute-force attacks. Moreover, in Figure~\ref{fig:cdf_zx} we plot the proportion of passwords with each zxcvbn score, from 0 to 4, for both datasets. It is clear that the developers' passwords dominate the higher scores (3 and 4), further confirming the enhanced security awareness among developers, who tend to create stronger and more secure passwords compared to normal users. Interestingly, for the lowest score (0), we also observe a higher fraction of developers' passwords compared to 
\texttt{RockYou}, albeit very small ($<1\%$). This can be attributed to the fact that several database servers, etc., do not enforce password complexity requirements, allowing for weak passwords to be used. On the contrary, most of the online services from which the passwords of the \texttt{RockYou} dataset were leaked have enforced stricter password policies for user accounts preventing users from having passwords with a very low score.

\begin{figure*}[th]
\centering
	\includegraphics[width=.9\textwidth]{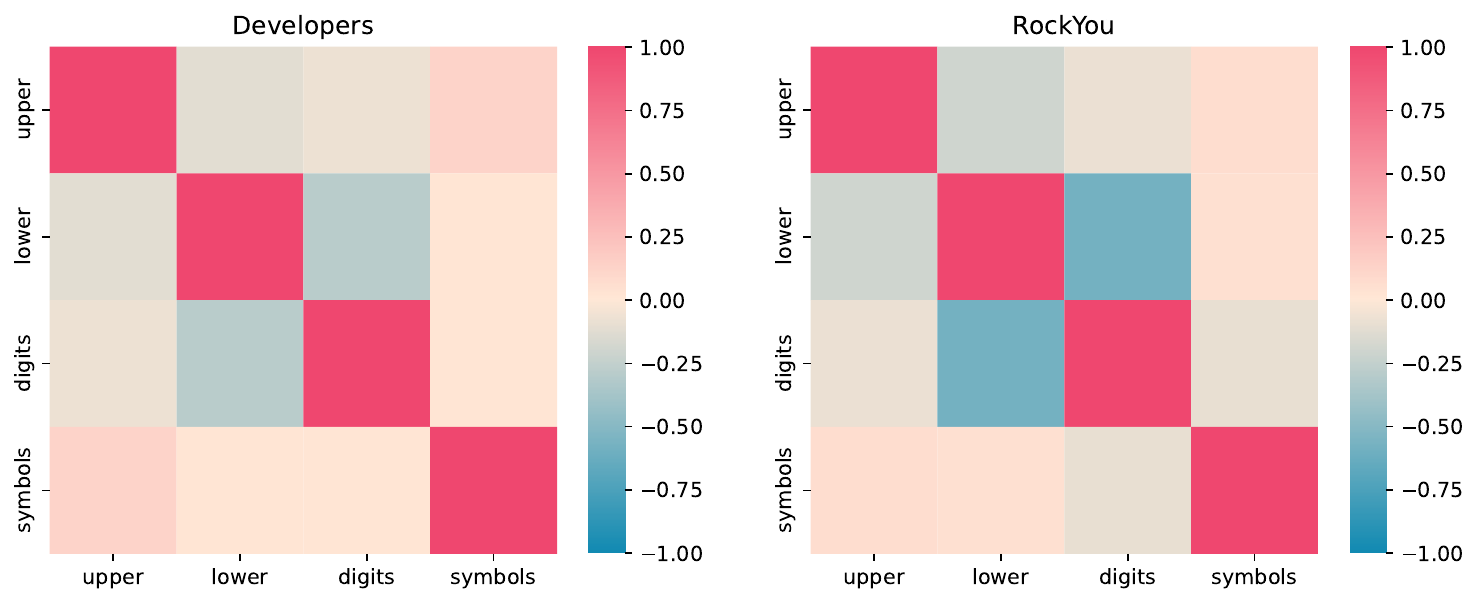}
    \caption{ASCII feature correlation matrices.}
    \label{fig:corr_matrix}
\end{figure*}

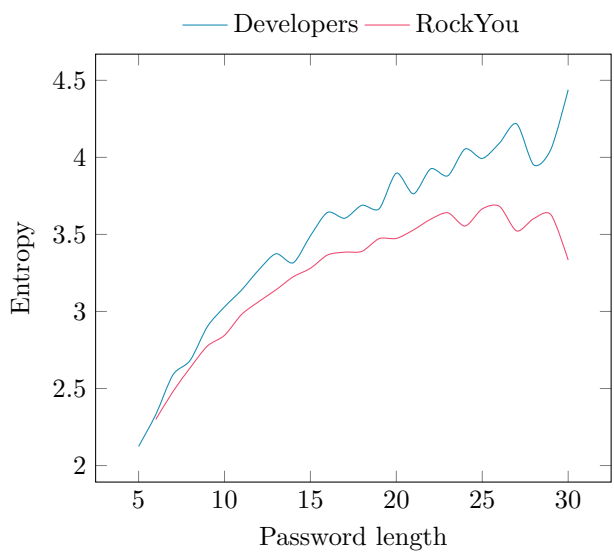
\begin{figure}[th]
\centering
\begin{tikzpicture}
\begin{axis}[
legend columns=-1,
        legend style={draw=none,at={(0.5,1.12)},anchor=north},
    xlabel=Password length,
    ylabel=Entropy,
    xtick={5,10,15,...,30},
            ]
\addplot[smooth,myblue] plot coordinates {
(5 ,    2.124155)
(6 ,    2.330366)
(7 ,    2.588620)
(8 ,    2.682399)
(9 ,    2.901056)
(10,    3.029707)
(11,    3.139917)
(12,    3.270605)
(13,    3.374542)
(14,    3.316119)
(15,    3.490997)
(16,    3.643305)
(17,    3.605189)
(18,    3.689042)
(19,    3.666440)
(20,    3.897710)
(21,    3.763562)
(22,    3.925550)
(23,    3.880688)
(24,    4.054554)
(25,    3.993146)
(26,    4.092543)
(27   , 4.218046)
(28  ,  3.950918)
(29 ,   4.053905)
(30,    4.438588)
};
\addlegendentry{Developers}

\addplot[smooth,color=myred]
    plot coordinates {
(6,     2.299019)
(7,     2.479743)
(8,     2.634353)
(9,     2.775185)
(10,    2.845170)
(11,    2.981077)
(12,    3.064381)
(13,    3.140988)
(14,    3.224937)
(15,    3.280141)
(16,    3.366732)
(17,    3.384924)
(18,    3.390079)
(19,    3.472848)
(20,    3.473959)
(21,    3.529157)
(22,    3.598839)
(23,    3.640371)
(24,    3.554571)
(25,    3.665790)
(26,    3.682426)
(27,    3.522508)
(28,    3.602894)
(29,    3.627496)
(30,    3.333927)
    };
\addlegendentry{RockYou}
\end{axis}
    \end{tikzpicture}
    \caption{Comparison of the entropy of the passwords per length.}
    \label{fig:entropy}
\end{figure}

In Figure \ref{fig:entropy} we compare the entropy of the passwords of the two datasets per password length. Clearly, the developers' passwords have more entropy than the passwords that typical users select for all lengths. Notably, this difference increases as the password length increases. Thus, we can again conclude that the developers' passwords are more random and thus more secure than the ones of users.

Finally, we focus on four key features of passwords: the number of uppercase ASCII characters, lowercase ASCII characters, digits, and symbols. These features have been widely studied in the context of password security and strength \cite{dinev2006extended,yan2004password}. A diverse combination of these character types is crucial to creating more secure and less predictable passwords, as it increases the search space for potential attacks and makes it more challenging for attackers to guess passwords using brute-force or dictionary-based methods. We present the pairwise correlation matrices (Pearson) of these features for the \texttt{Developers} and \texttt{RockYou} datasets in Figure~\ref{fig:corr_matrix}. We observe a significant difference in the correlation patterns between the two classes of passwords. For the Developers class, the correlation between uppercase and lowercase characters is 0.158, while for the RockYou class, it is -0.261. The positive correlation suggests that developers tend to incorporate a better mix of uppercase and lowercase letters in their passwords. Furthermore, the correlation between uppercase characters and symbols in \texttt{Developers} is 0.120, compared to only 0.027 for \texttt{RockYou}. This indicates that developers are more likely to include a combination of uppercase letters and symbols, another factor contributing to stronger passwords. On the other hand, the correlation between lowercase letters and digits in \texttt{Developers} is -0.134, whereas it is -0.652 for \texttt{RockYou}. This negative correlation in the RockYou class implies that as the number of lowercase letters increases, the number of digits decreases, leading to a less diverse and more predictable password composition. In contrast, the weaker negative correlation in the Developer class suggests a more balanced distribution of lowercase letters and digits. Based on the above, we conclude that the stronger cross-category correlations observed in developer passwords imply that they are more likely to contain a diverse mix of character types. As a result, the developers' passwords are more secure than the ones of normal users.

\section{Conclusions}
While shifting from DevOps to DevSecOps, many concepts, tools, methods, and pipelines have to be revised to prevent vulnerabilities from reaching final products and services. Password leaks from source code are a common security issue that significantly impacts numerous organisations world wide. In this work we perform the first analysis of real-world developers' passwords leveraging a large scale dataset that we collected from public GitHub repositories.

Additionally, we compare them to leaked user passwords from the RockYou2021 dataset. Our findings highlight that developers generally exhibit stronger password practices, with longer and more complex passwords. However, we also observed cases where developers' passwords were weak, particularly when certain systems did not enforce password complexity requirements. These results emphasise the need for continued education and awareness about secure password practices among developers, as well as the importance of enforcing password policies across all types of systems and services. Of course, this also underlines the extensive carelessness of developers which do not consider the risks that they expose their clients when committing their code on public repositories, acting as if their code is hosted in their own premises and not accessible from anyone else.  

\section*{Acknowledgements}
This work was supported by the European Commission under the Horizon Europe Programme, as part of the project LAZARUS (\url{https://lazarus-he.eu/}) (Grant Agreement no. 101070303).

The content of this article does not reflect the official opinion of the European Union. Responsibility for the information and views expressed therein lies entirely with the authors.

\end{document}